\documentclass[aip, jap, graphicx, reprint, floatfix]{revtex4-1}
\usepackage{graphicx}
\usepackage{dcolumn}
\usepackage{bm}

\usepackage[utf8]{inputenc}
\usepackage[T1]{fontenc}
\usepackage[english]{babel}
\usepackage{amsthm, amsmath, amssymb, tipa, graphicx, caption, subcaption, float, fancyhdr, graphics, verbatim, cancel, textgreek, etoolbox, mathptmx}

\draft 

\begin{document}


\title{Strong Negative Electrothermal Feedback in Thermal Kinetic Inductance Detectors}

\author{Shubh Agrawal}
\email{shubh@caltech.edu}
\affiliation{Department of Physics, California Institute of Technology, Pasadena, CA, 91125, USA}

\author{Bryan Steinbach}
\email{bsteinba@caltech.edu}
\affiliation{Department of Physics, California Institute of Technology, Pasadena, CA, 91125, USA}

\author{James J. Bock}
\affiliation{Department of Physics, California Institute of Technology, Pasadena, CA, 91125, USA}
\affiliation{Jet Propulsion Lab, Pasadena, CA, 91109, USA}
\author{Clifford Frez}
\affiliation{Jet Propulsion Lab, Pasadena, CA, 91109, USA}
\author{Lorenzo Minutolo}
\affiliation{Department of Physics, California Institute of Technology, Pasadena, CA, 91125, USA}
\author{Hien Nguyen}
\affiliation{Jet Propulsion Lab, Pasadena, CA, 91109, USA}
\author{Roger O'Brient}
\affiliation{Jet Propulsion Lab, Pasadena, CA, 91109, USA}
\author{Anthony Turner}
\affiliation{Jet Propulsion Lab, Pasadena, CA, 91109, USA}
\author{Albert Wandui}
\affiliation{Department of Physics, California Institute of Technology, Pasadena, CA, 91125, USA}

\date{\today}

\begin{abstract}
We demonstrate strong negative electrothermal feedback accelerating and linearizing the response of a thermal kinetic inductance detector (TKID). TKIDs are a proposed highly multiplexable replacement to transition-edge sensors and measure power through the temperature-dependent resonant frequency of a superconducting microresonator bolometer. At high readout probe power and probe frequency detuned from the TKID resonant frequency, we observe electrothermal feedback loop gain up to $\mathcal L$ $\approx$ 16 through measuring the reduction of settling time. We also show that the detector response has no detectable non-linearity over a 38\% range of incident power and that the noise-equivalent power is below the design photon noise.
\end{abstract}


\maketitle 

\section{Introduction And Motivation}

We present observations of strong negative electrothermal feedback in a thermal kinetic inductance detector (TKID). TKIDs are cryogenic bolometers that detect minute power fluctuations by measuring the temperature fluctuations of a suspended absorber \cite{ulbricht_highly_2015, quaranta_x-ray_2013, arndt_optimization_2017, timofeev_submillimeter-wave_2014, dabironezare_dual_2018, wernis_characterizing_2013}. The suspended absorber is connected to a thermal bath with a weak thermal link so that the incident power and the suspended absorber temperature are related approximately linearly. In a TKID, the temperature rise is measured through the temperature dependence of the kinetic inductance effect in a superconducting inductor on the suspended absorber. The inductor is coupled to a capacitor to form a superconducting microresonator, such that the incident power is measured by the change in the resonant frequency. The resonant frequency is measured through the phase shift of a readout probe signal, which normally is at a low enough power (much less than the incident power which is being measured) such that the TKID dynamics are not altered by the probe.

Negative electrothermal feedback occurs in bolometers when the power dissipated in the temperature sensor has a negative temperature dependence. Strong electrothermal feedback through Joule heating reduces non-linearities and resolution limitations in voltage-biased transition-edge sensors \cite{irwin_application_1995}. Feedback reduces Johnson noise in bolometers using resistive sensors \cite{mather_bolometer_1982}. In a detector with negative electrothermal feedback, dissipated readout power decreases rapidly with increasing temperature, such that temperature deviations from the operating point return to the mean more rapidly. Total power flux is held nearly constant, as readout power compensates for changes in incident power, which increases the linearity of the bolometer. In the strong electrothermal feedback regime with transition-edge bolometers, this effect reduces the time constant of the detector by an order of magnitude \cite{irwin_transition-edge_2005}. These benefits have led to numerous applications for transition-edge sensors, where strong electrothermal feedback produces fast, linear, and photon noise limited sensors for millimeter-wave detection \cite{anderson_performance_2020, zhang_characterizing_2020, koopman_advanced_2018} and X-ray calorimeters with eV energy resolution \cite{szypryt_transition-edge_2019}.

\begin{figure}[!ht]
\includegraphics[width=\columnwidth]{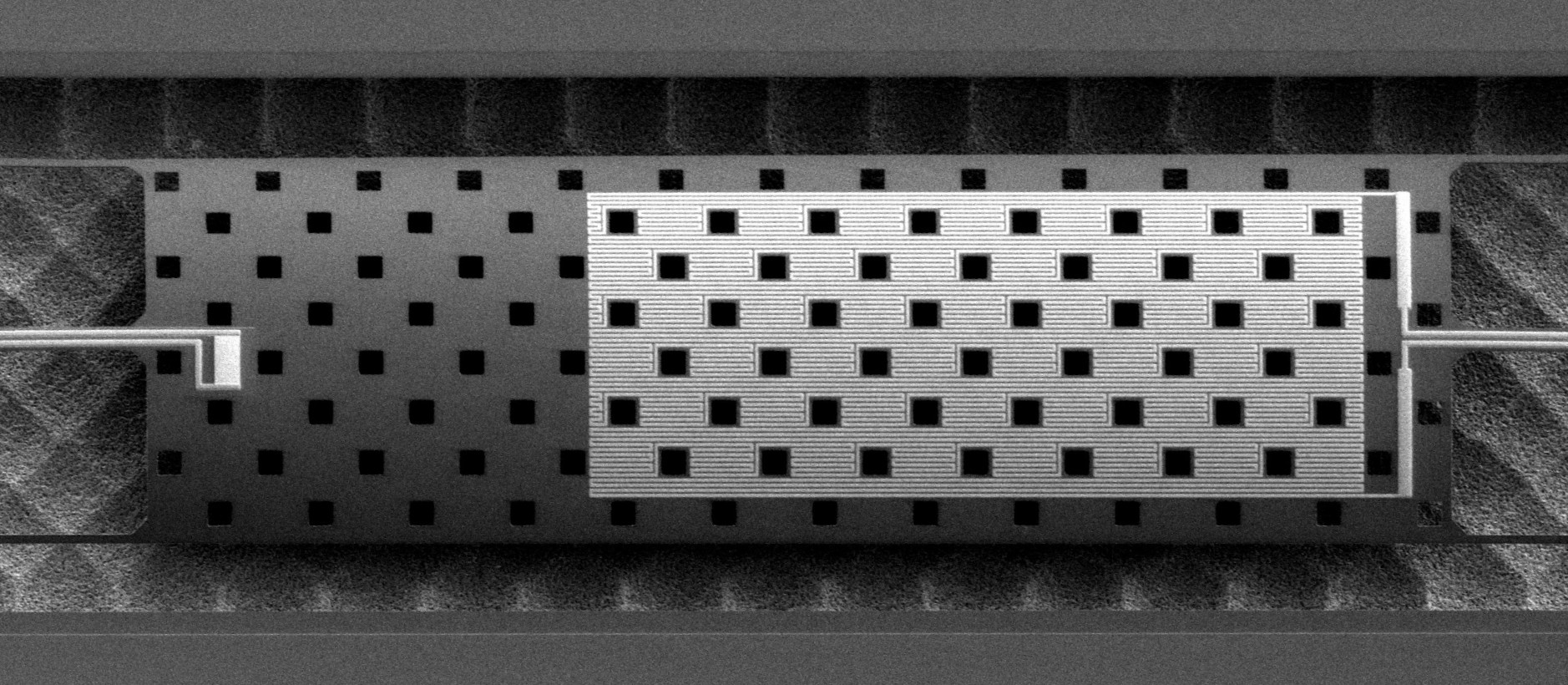}
\caption{\label{fig:bolo} SEM micrograph of the suspended absorber of a TKID of the design used for this work. The gold resistor heater is seen on the left and the meandered aluminum inductor is seen on the right. For more details on the design of the TKID, see Wandui et al. \protect{\cite{wandui_thermal_2020}}}
\end{figure}

Lindeman \cite{lindeman_resonator-bolometer_2014} proposed a mechanism through which electrothermal feedback would occur in a TKID when the frequency of the readout probe signal was detuned from the resonant frequency. Detuning the probe frequency above the TKID's resonant frequency, an increase in temperature of the suspended absorber decreases the resonant frequency due to an increase in kinetic inductance, moving the resonant frequency further from the probe frequency. This decreases the electrical power dissipated by the probe in the resonator, which in turn decreases the temperature of the suspended absorber, resulting in negative electrothermal feedback. The feedback can be strong in resonators with high quality factors, as there is a large change in absorption of power from the probe signal for a small change in resonant frequency. We demonstrate that this feedback occurs in a TKID device of the design previously presented and characterized at low readout power in Wandui et al \cite{wandui_thermal_2020} and shown in Fig. \ref{fig:bolo}.

\subsection{\label{sec:model} Modeling}

\begin{figure}[!ht]
\includegraphics[width=\columnwidth]{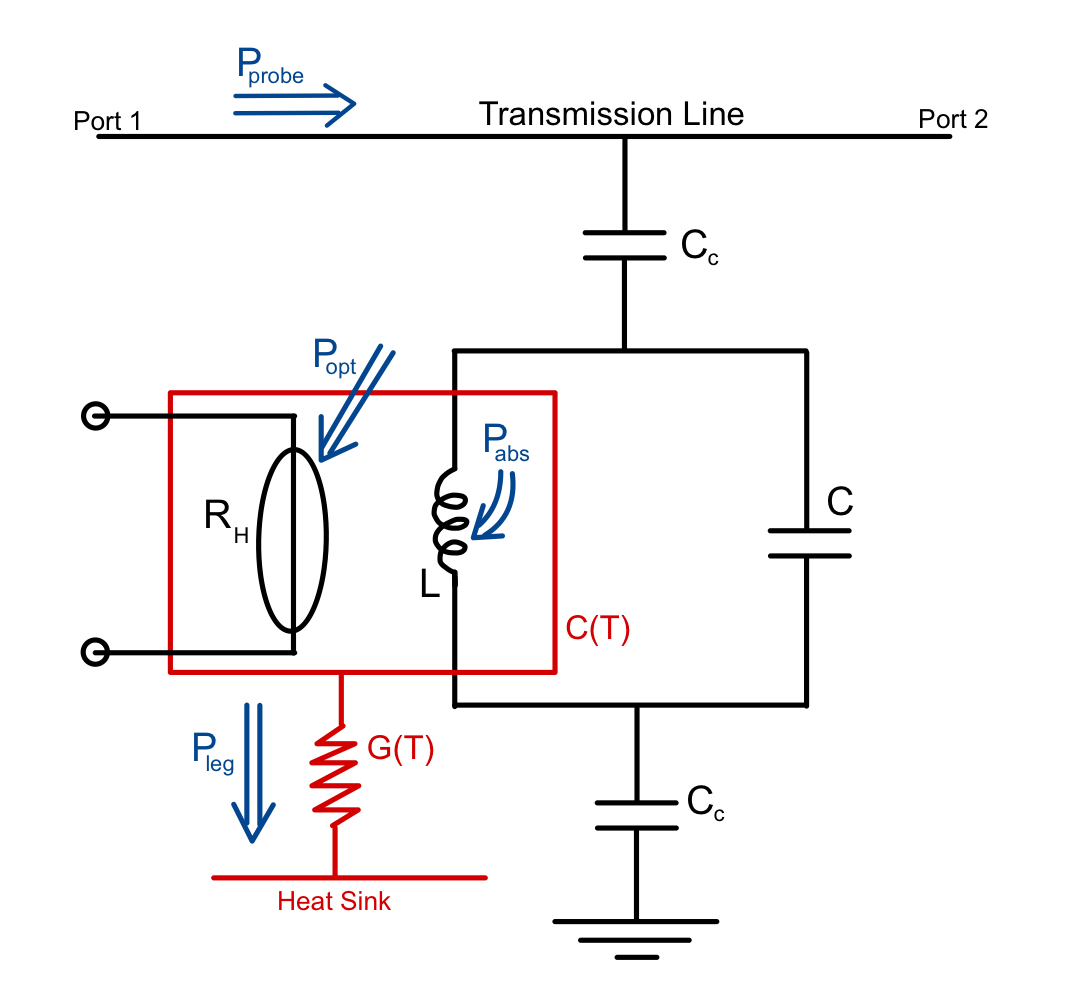}
\caption{\label{fig:circuit} Schematic and circuit diagram of the TKID. The suspended absorber, as shown in Fig. \ref{fig:bolo}, is enclosed in red. The blue arrows represent the power fluxes into ($P_{opt}$ and $P_{abs}$) and out of ($P_{leg}$) the absorber and the probe power along the transmission line ($P_{probe}$). Capacitors $C_c$ weakly couple the detector to the transmission line. The inductance L and capacitance C form the TKID's superconducting microresonator.}
\end{figure}

We model the response of a TKID to probe signals of high power following the schematic in Fig. \ref{fig:circuit}. Probe power $P_{probe}$ enters port 1 of the transmission line and is modulated by the resonant circuit, and some phase-shifted fraction of it exits at port 2.
Strong electrothermal feedback manifests as a non-linear response to high probe powers reminiscent of the Duffing oscillator. Similar behavior is observed in traditional hot quasiparticle kinetic inductance detectors due to the non-linearity of kinetic inductance near the superconducting critical current \cite{swenson_operation_2013}. To determine the non-linear behavior of a TKID, we first solve for the stable non-linear operating temperatures of the suspended absorber and then calculate the strength of the electrothermal feedback from the relation between probe power dissipation and absorber temperature.

Incident power $P_{opt}$ plus the readout probe power dissipated in the resonator $P_{abs}$ heat the suspended absorber. $P_{abs}$ is a fraction of the probe power $P_{probe}$, 

\begin{equation}
\label{eqn:pabs}
   P_{abs} = \frac{1}{2} \chi_c \chi_g P_{probe}.
\end{equation}

The dissipated probe power is limited by the coupling efficiency $\chi_c$ of the resonator to the transmission line, and the detuning efficiency $\chi_g$ of the resonator to the probe frequency. We generalize $\chi_c$ to complex coupling quality factors $Q_c$ with complex angle $\phi_c$ for asymmetric resonances \cite{khalil_analysis_2012}, where $x = (f-f_r)/f_r$ is the fractional frequency detuning and $Q_i$ and $Q_r$ are the internal and total quality factors,

\begin{equation}
\label{eqn:chi}
\chi_c = \frac{4 |Q_c| Q_i}{(|Q_c| + Q_i \cos(\phi_c)^2)}, \quad \chi_g = \frac{1}{1 + 4 Q_r^2 x^2}.
\end{equation}

The fractional energy loss per cycle from the resonator $Q_r^{-1}$ is a reciprocal sum of the absorption in the inductor $Q_i^{-1}$ and loss to the transmission line $\Re \left[ Q_c^{-1} \right]$,

\begin{equation} 
\label{eqn:Qr}
Q_r^{-1} = Q_i^{-1} + \Re \left[ Q_c^{-1} \right].
\end{equation}

At low temperature, the resonant frequency is $f_0$. As the temperature rises, the frequency shifts to $f_r$ due to the temperature dependence of the kinetic inductance effect. We assume the superconductor follows Mattis-Bardeen theory for a thin film in its frequency shift $x_{MB} = (f_r-f_0)/f_0$ and internal quality factor $Q_{i}$,

\begin{equation}
\label{eqn:xshift_qr}
x_{MB} = - \frac{\alpha_k S_2 n_{qp}}{4 N_0 \Delta},
\end{equation}

\begin{equation}
    \label{eqn:Qi}
    Q_{i}^{-1} = \frac{2 N_0 \Delta}{\alpha_k n_{qp} S_1},
\end{equation}
where $S_1$ and $S_2$ are the Mattis-Bardeen derived absorption and frequency responses for a superconducting microresonator \cite{zmuidzinas_superconducting_2012}. $n_{qp}$ is the equilibrium quasiparticle density for a BCS superconductor at suspended absorber temperature $T$ with gap $\Delta$ and density of states $N_0$. For our aluminum films with critical temperature $T_c$, we use $\Delta$ $\approx$ 1.75 $k_B T_c$ = 0.00015 eV/K $\times$ $T_c$ and $N_0 = 1.72 \times 10^{10}$ \textmu{}m$^{-3}$ eV$^{-1}$. We use low temperature approximations to the Mattis-Bardeen integrals for $n_{qp}$, $S_2$ and $S_1$,

\begin{equation}
\label{eqn:nqp}
n_{qp} = 2 N_0 \sqrt{2\pi k_B T \Delta} \exp\Big(-\frac{\Delta}{k_B T}\Big),
\end{equation}
\begin{equation}
\label{eqn:s2}
S_2 = 1 + \sqrt{\frac{2 \Delta}{\pi k_B T}} \exp\Big(- \frac{hf}{2k_B T}\Big) I_0 \Big[ \frac{hf}{2k_B T} \Big],
\end{equation}
\begin{equation}
\label{eqn:s1}
S_1 = \frac{2}{\pi} \sqrt{\frac{2 \Delta}{\pi k_B T}} \sinh\Big(\frac{hf}{2k_B T}\Big) K_0 \Big[ \frac{hf}{2k_B T} \Big].
\end{equation}

The suspended absorber has a heat capacity $C(T)$ and is connected to a heat sink at temperature $T_{bath}$ through a thermal conductance $G(T)$ = $n K_c T^{n-1}$ = $\partial P_{leg} / \partial T$. The power transferred between the suspended absorber and the heat sink is $P_{leg}$ = $K_c (T^n - T_{\text{bath}}^n)$. 

The thermal energy in the suspended absorber changes at a rate equal to net power entering and leaving the suspended absorber,

\begin{equation}
\label{eqn:heat}
C(T) \frac{d T}{d t} = - P_{leg} + P_{abs} + P_{opt}.
\end{equation}

\begin{figure}[!ht]
\includegraphics[width=\columnwidth]{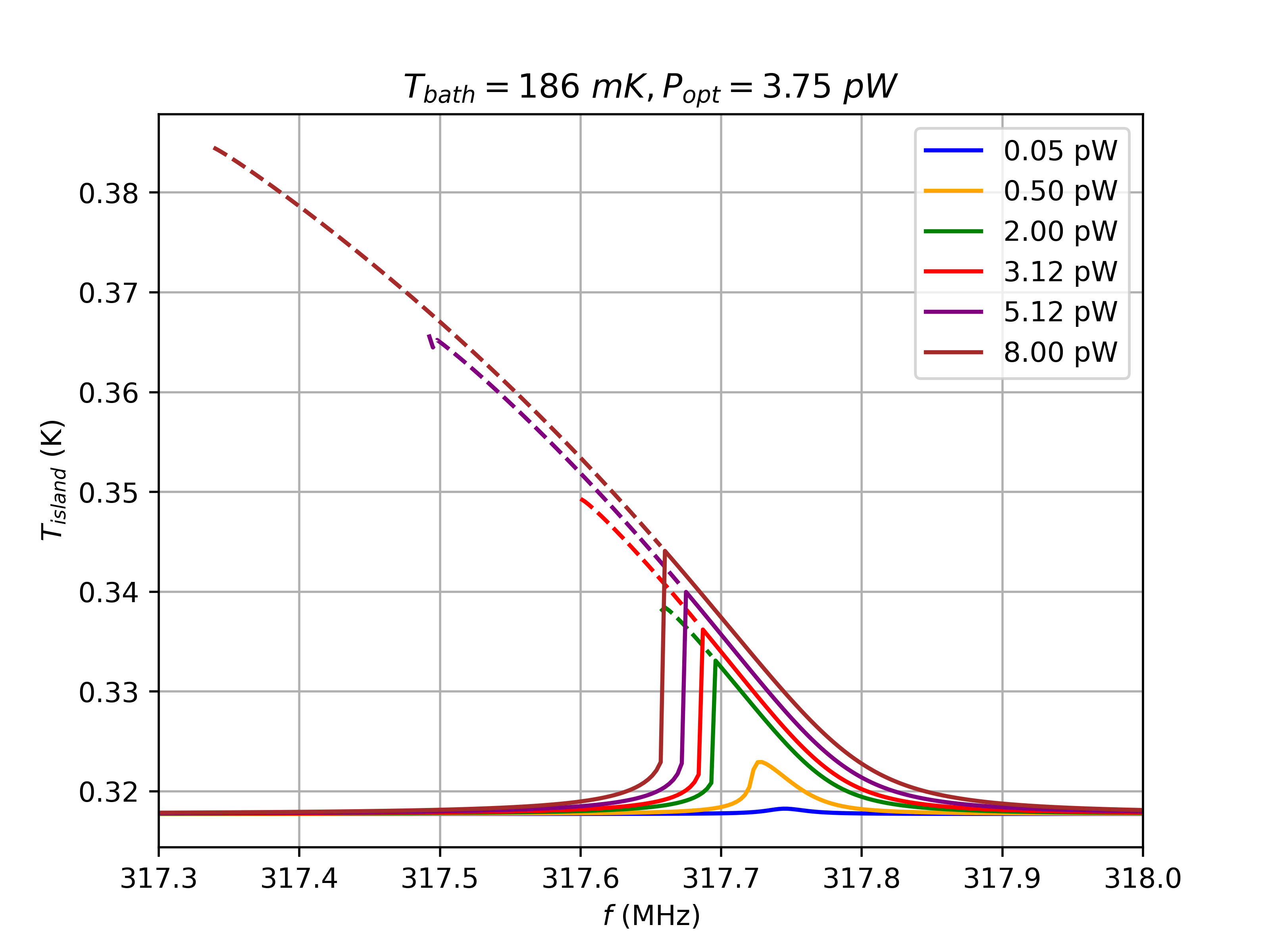}
\caption{\label{fig:curves} Model predictions for suspended absorber temperature $T$ under slow probe frequency sweeps. Different probe powers are shown in different colors. Solid lines indicate the cold branch which is accessed through frequency up sweeps; dashed segments show the hot branch which is accessed through frequency down sweeps. Model parameters used are those for the device tested in Section \ref{character}.}
\end{figure}

The steady-state solutions for the temperature thus satisfy $P_{abs}(T)$ + $P_{opt}$ = $P_{leg}(T)$. In order to be a stable solution, we additionally require $ \partial (P_{abs} - P_{leg}) / \partial T < 0$, as otherwise a small perturbation in temperature will grow exponentially away from equilibrium. We show predicted stable temperature solutions for a range of probe powers in Fig. \ref{fig:curves}.

Like the Duffing oscillator, at high probe powers, the response to probe frequency sweeps splits into two stable branches. The stable branches are a cold branch with weak positive electrothermal feedback and a hot branch with strong negative electrothermal feedback. A third unstable branch exists at an intermediate temperature and experiences strong positive electrothermal feedback, so small perturbations grow rapidly away from equilibrium.  We neglect this unstable solution, which is inaccessible to our experimental setup.

Qualitatively, we explain the extended hot branch and the lower bifurcation frequency in Fig. \ref{fig:curves} as follows. If the probe frequency starts above the resonant frequency, we dissipate a small amount of power in the resonator that pushes its frequency slightly down. As we lower the probe frequency and have it approach the resonant frequency, the dissipated power grows, so the resonant frequency is further pushed down. The frequency gap must continue to close for the dissipated power to grow, so $\chi_g$ increases. Eventually, the frequency gap shrinks to zero, and the dissipated power is maximized. At this point, any further decrease in probe frequency decreases the dissipated power, and the resonant frequency begins rising and bifurcates back to the cold branch, leaving the probe frequency well below the resonant frequency. 

We term these sweeps described above, where the probe frequency is stepped down slowly in small decrements, as ``down sweeps''. ``Up sweeps'' are where we step the probe frequency up slowly in small increments, and these exhibit positive electrothermal feedback. In up sweeps, the probe frequency starts lower than the resonance frequency, and as power is dissipated in the resonator and positive electrothermal feedback becomes significant, the resonant frequency is pulled down and bifurcates to below the probe frequency.

Non-linear kinetic inductance \cite{swenson_operation_2013} and quasiparticle heating \cite{thomas_electrothermal_2015, guruswamy_electrothermal_2017} could produce similar non-linearities in our devices, without producing useful electrothermal feedback that speeds up the time constant or linearizes the response. In the case of non-linear kinetic inductance, as the current in the inductor approaches the critical current, the inductance increases, and the resonant frequency decreases, mimicking electrothermal feedback from the perspective of the probe. We minimize the impact of non-linear kinetic inductance in our setup by operating the TKID at higher temperatures, where it has strong responsivity, and therefore relatively low quality factors ($\approx$ 10,000) and low current densities in the superconducting film.

Quasiparticle heating can also produce electrothermal feedback internal to the quasiparticle population, but without any useful reduction in the time constant. Quasiparticle heating is reduced due to our large volume inductor, which maximizes the thermal conductance between the quasiparticles and the phonons in the superconducting film. The quasiparticle to phonon thermal conductance also benefits from high temperatures, where the quasiparticle density is high and the quasiparticle lifetime is short.

\subsection{Loop Gain and Time Constant}

We model the dynamics of small perturbations in temperature by linearizing Eq. \ref{eqn:heat}. We Taylor expand $P_{leg} = P_{leg}^0 + G T$ where $G = (\partial P_{leg})/T$, $P_{abs} = P_{abs}^0 - G_{ETF} T$, where $G_{ETF} = (\partial P_{abs})/{T}$, and where $T$ now refers to a small perturbation in temperature from the equilibrium. With these substitutions the linearized version of Eq. \ref{eqn:heat} becomes

\begin{equation}
\label{eqn:bolometeretf}
C \frac{d T}{d t} = -(G + G_{ETF}) T + P_{opt}(t).
\end{equation}

Following the convention for transition-edge sensors \cite{irwin_transition-edge_2005}, we define loop gain as $\mathcal L = G_{ETF}/G$. Loop gain measures how strongly electrothermal feedback forces the temperature of the bolometer back to equilibrium, relative to the forcing due to the thermal link. Negative electrothermal feedback corresponds to positive loop gain due to the minus sign in the definition of $G_{ETF}$. Solving the $P_{opt}(t)$ step response of Eq. \ref{eqn:bolometeretf} leads to the time constant for the bolometer

\begin{equation} 
\label{eqn:loopgain}
\tau_{ETF} = \frac{1}{\mathcal L + 1} \frac{C}{G} = \frac{1}{\mathcal L + 1} \tau_0,
\end{equation}
where $\tau_0 = C/G$ is the intrinsic bolometer thermal time constant.

The effective electrothermal feedback $G_{ETF}$ is the derivative of $P_{abs}$ with temperature, which from Eq. \ref{eqn:pabs} is

\begin{equation}
G_{ETF} = P_{abs}'(T) = \frac{1}{2} (\chi_c'(T) \chi_g(T) + \chi_c(T) \chi_g'(T)) P_{probe}.
\end{equation}

To estimate loop gain in the strong negative electrothermal feedback regime, the term $\chi_c'(T)$ can be neglected, because it is temperature-dependent only through the internal quality factor, while $\chi_g'(T)$ is temperature-dependent through resonator frequency shifts. At the low probe frequencies of interest for TKIDs, the frequency shift effects are more significant than quality factor shifts by typically about an order of magnitude. We additionally assume that $Q_r'(T)$ is zero in $\chi_g'(T)$. Then, the temperature dependence of $G_{ETF}$ is contained in

\begin{equation}
\chi_g'(T) = -8 Q_r^2 \chi_g(T)^2 x(T) x'(T).
\end{equation}

The fractional frequency offset between probe and resonator that maximizes $\chi_g'(T)$ is 

\begin{equation}
\hat x = \frac{1}{Q_r \sqrt{12}}.
\end{equation}

At this detuning (where the probe is between one-quarter and one-third of a linewidth above the resonant frequency), the power absorbed is

\begin{equation}
\hat P_{abs} =\frac{3}{8} \chi_c P_{probe}.
\end{equation}

The loop gain at optimal detuning $ \hat {\mathcal L} $ can be expressed in terms of $ \beta = S_2/S_1 $, $ \kappa = \partial \log n_{qp} \Big/ \partial \log T $. We additionally define $\alpha = \sqrt{3} \beta \kappa / 4$, in order to put the expression of the loop gain of a TKID into the same form as the loop gain of a transition-edge sensor\cite{irwin_transition-edge_2005}:

\begin{equation}
\label{eqn:maxloopgain}
    \hat {\mathcal L} = \frac{\alpha P_{abs}}{G T}.
\end{equation}

\subsection{Noise Model\label{sec:noisemodel}}
The intrinsic noise of a TKID contains contributions from three sources: thermal fluctuations in the suspended absorber due to the exchange of phonons in the weak thermal link, quasiparticle number density fluctuations in the superconductor due to thermal generation and recombination of quasiparticles, and readout noise from the low noise amplifier \cite{wandui_thermal_2020}. When the TKID is operated in the strong electrothermal feedback regime, we expect a significant change in the phonon noise and small changes in generation recombination noise and readout noise. We expect our devices to be strongly phonon noise limited. The phonon noise contribution to the noise-equivalent power of the detector, 

\begin{equation}
\label{eqn:nepphonon} NEP_{\text{phonon}} = \sqrt{4 k_B T^2 G \gamma} ,
\end{equation}
includes $\gamma \approx 0.5$, a factor that accounts for the lower temperature of the thermal bath\cite{mather_bolometer_1982}. Phonon noise will generally increase due to an increase in the total thermal power in the legs. A typical operating condition for transition-edge sensors is setting dissipated readout power equal to the incident power; for this setup, the phonon noise increases by a factor of $\sqrt{2}$ if the detector thermal conductance $G$ is re-optimized to fix the suspended absorber temperature.

Changes in the quasiparticle generation recombination noise should be limited by the large volume of our inductor, as the quasiparticle population is in thermal equilibrium with the phonons in the suspended absorber.
The readout noise should not vary much, because the decrease in coupling efficiency from detuning is compensated for by the increase in readout power, such that the absolute responsivity $P_{probe}$ $\times$ ${d S_{21}} / {d P_{opt}}$ is only weakly dependent on probe power in the strong electrothermal feedback regime.

We also expect an additional source of frequency-dependent noise due to finite thermal conductance within the suspended absorber and distributed heat capacity. This effect is observed in transition-edge sensors at high loop gains \cite{lee_voltage-biased_1998}, manifesting as a rise in noise beyond the intrinsic thermal bandwidth of the suspended absorber. Our suspended absorber is approximately 300 nm thick, much thinner than the 1000 nm thick absorbers where we have observed this effect in transition-edge sensors \cite{kernasovskiy_measuring_2020}. This reduces the thermal conductance within the absorber which is limited by the cross-sectional area of the suspended absorber, due to the phonon analog of the Stefan-Boltzmann law.

\section{\label{results} Results and Discussion}

We studied a single TKID at high readout probe powers using the cryostat described in Wandui et al \cite{wandui_thermal_2020} and the software-defined radio system described in Minutolo et al \cite{minutolo_flexible_2019}. To deposit incident power $P_{opt}$ that is free of photon noise, we supplied DC to the gold heater resistor visible in Fig. \ref{fig:bolo} on the suspended absorber. The nominal incident power used in our experiments is $P_{opt}$ = 3.75 pW which brings the aluminum inductor to a temperature $T\approx 0.33 K$ where it has useful responsivity.

\subsection{\label{character} TKID Characterization}

To characterize the TKID resonator at low readout probe power, we obtained transmission $S_{21}$ measurements by biasing the detector at a fixed probe power $P_{probe}$ = $0.1$ pW and sweeping the probe frequency $f$ over 2001 equally-spaced points between $312.5$ MHz and $319$ MHz. We repeated this for 10 different values of bath temperature $T_{bath}$ between 114 mK and 439 mK. A least-squares fit of the $S_{21}$ curves to Eq. \ref{eqn:s21} gives us the resonant frequency $f_r$ dependent on $T_{bath}$, which when fitted using Eqs. \ref{eqn:xshift_qr}, \ref{eqn:nqp}, and \ref{eqn:s2},  estimates $|Q_c|$ $\approx$ 13500, $\phi_{c}$ $\approx$ 0.3, $T_c$ = 1.32 $\pm$ 0.01 K, and $\alpha_k$ = 0.50 $\pm$ 0.03.

With the resonator now calibrated as a thermometer, we characterize the weak thermal link by sweeping the incident power $P_{opt}$ in 31 steps between 0 and 23.4 pW with $T_{bath} =$ 186 mK. Least-squares fits to the responses determines the dependence of $f_r$ on $P_{opt}$. The approximation $P_{opt}$ $\approx$ $P_{leg}$ in Eqs. \ref{eqn:heat} and \ref{eqn:pabs} gave us $f_0$ = 317.889 $\pm$ 0.005 MHz, $K_c$ = 185 $\pm$ 3 pW / K$^n$, and $n$ = 3.23 $\pm$ 0.02. Our value of $n \approx 3$ is consistent with similar bolometers with thin silicon nitride support legs used in BICEP/Keck \cite{wandui_thermal_2020}.

\subsection{\label{s21} Non-linear $S_{21}$ curves}

We measured non-linear $S_{21}$ by biasing the detector at a range of fixed probe powers $P_{probe}$ and performing both up and down sweeps, over 3000 equally-spaced probe frequencies $f$ between $316$ MHz and $318.4$ MHz. The bath temperature was regulated to 186 mK, and we fixed the incident power at $P_{opt} =$ 3.75 pW. The predicted $S_{21}$ is given by

\begin{equation}
\label{eqn:s21}
S_{21}(f) = 1 - \frac{Q_r}{Q_c} \frac{1}{1 + j 2 Q_r x},
\end{equation}
where $Q_r$ and $x$ are determined by solving for the stable suspended absorber temperature as described in Section \ref{sec:model}.

\begin{figure}[!ht]
\begin{subfigure}{\columnwidth}
\includegraphics[width=\columnwidth]{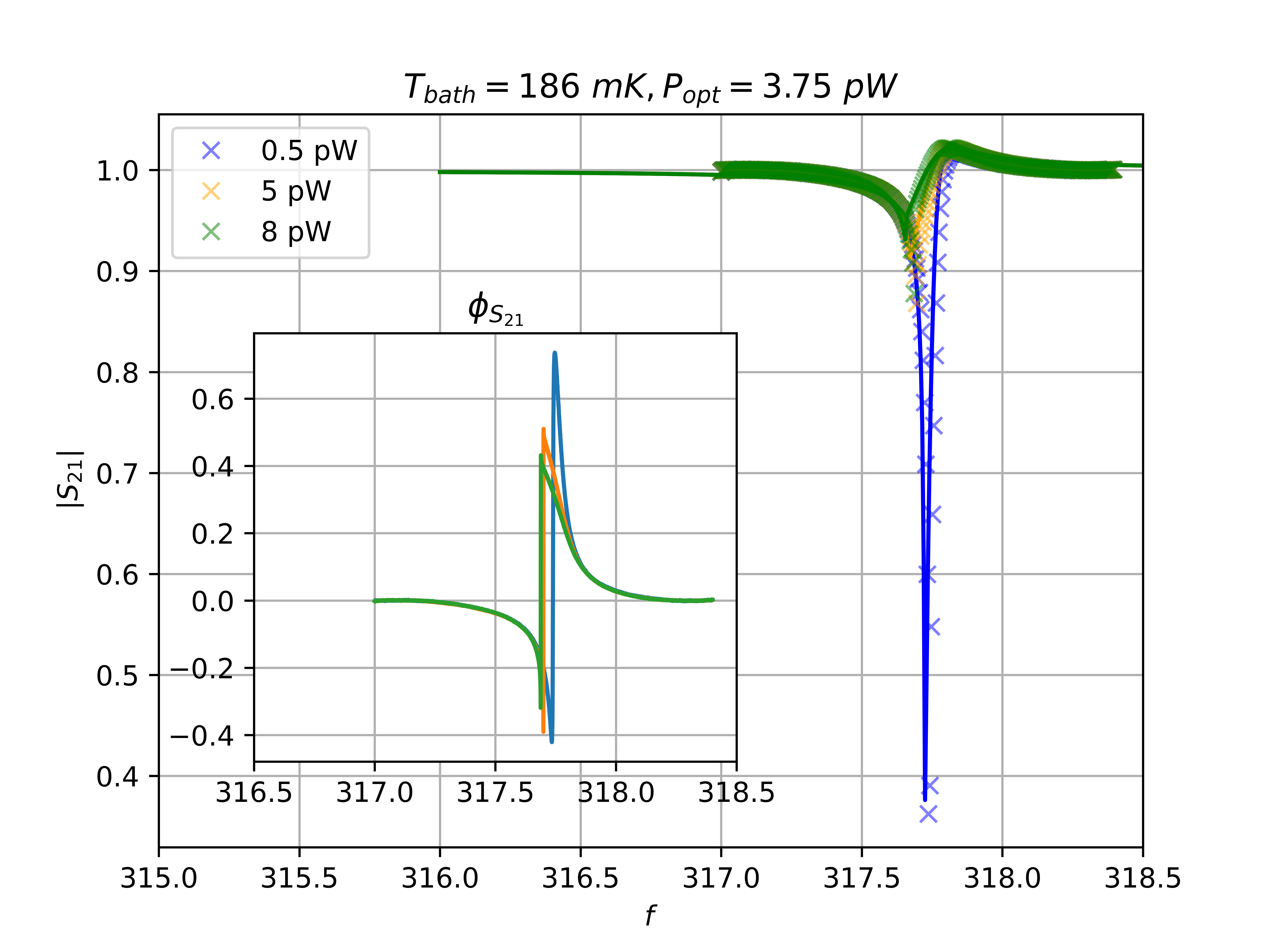}
\caption{Up Sweeps: Frequency stepped up in gradual increments.}
\end{subfigure}
\begin{subfigure}{\columnwidth}
\includegraphics[width=\columnwidth]{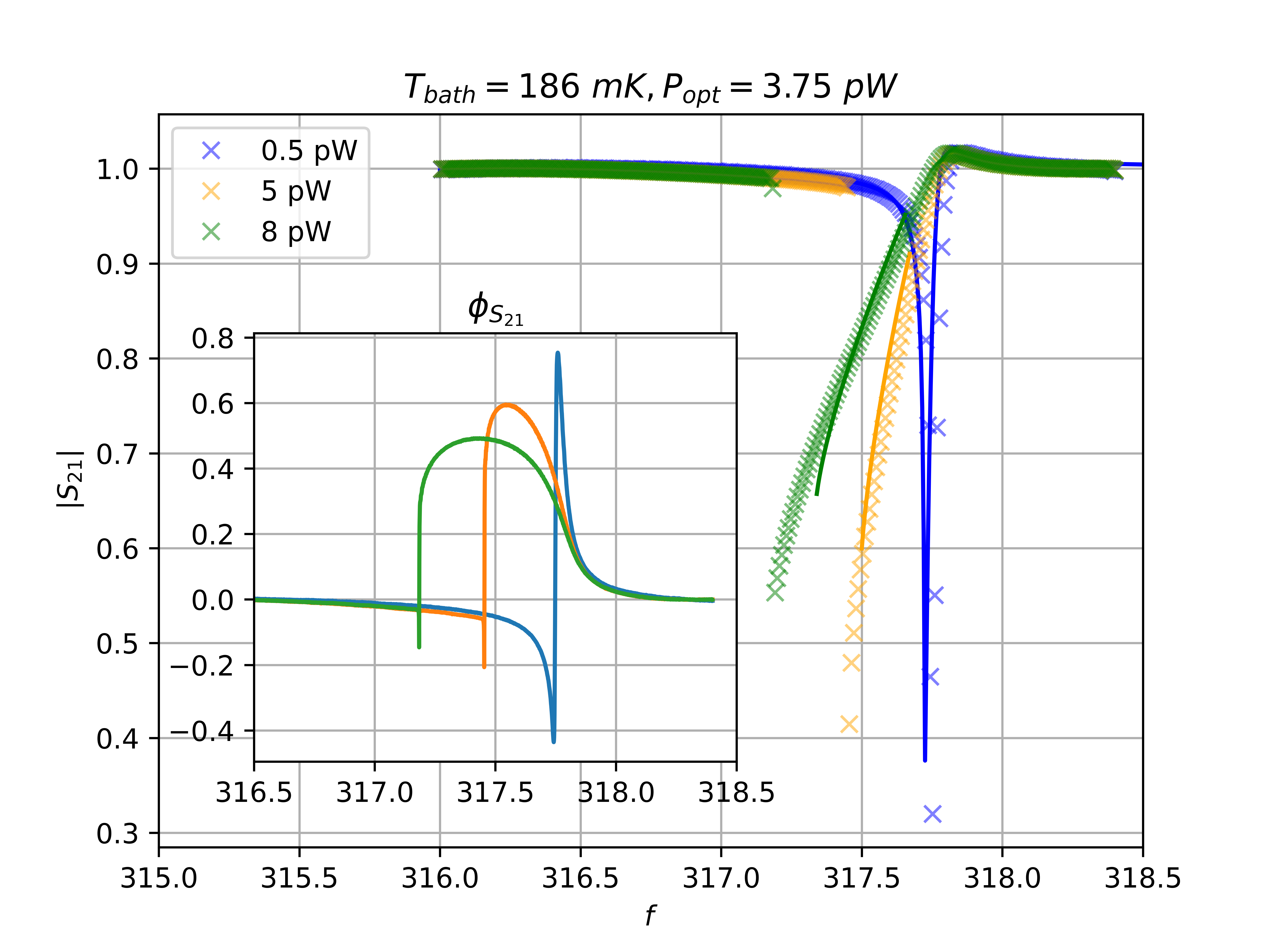}
\caption{Down Sweeps: Frequency stepped down in gradual decrements.}
\end{subfigure}
\caption{\label{fig:S21sweeps}  Crosses show the measured $S_{21}$ magnitude plotted against frequency; model predictions are given by lines. Insets show measured $S_{21}$ phase. Strong hysteresis is seen at probe powers of the order of a few pW, when $P_{probe}$ is comparable to $P_{opt}$.}
\end{figure}

Fig. \ref{fig:S21sweeps} shows the measured and predicted $S_{21}$ frequency sweeps when probe power at the chip is $0.5$ pW, $5$ pW, and $8$ pW. Directional hysteresis is absent at 0.5 pW but shows up strongly at 5 and 8 pW. For frequency down sweeps at the higher probe powers, the resonance frequency is pushed down as predicted by the negative electrothermal feedback model. At the highest probe power setting with $P_{probe}$ = 8 pW, we observed that the lower bifurcation frequency of the hysteresis branch deviates by a few MHz from the resonance frequency obtained at low probe power.

\subsection{\label{tau} Speed of Response}

\begin{figure}[!ht]
\includegraphics[width=\columnwidth]{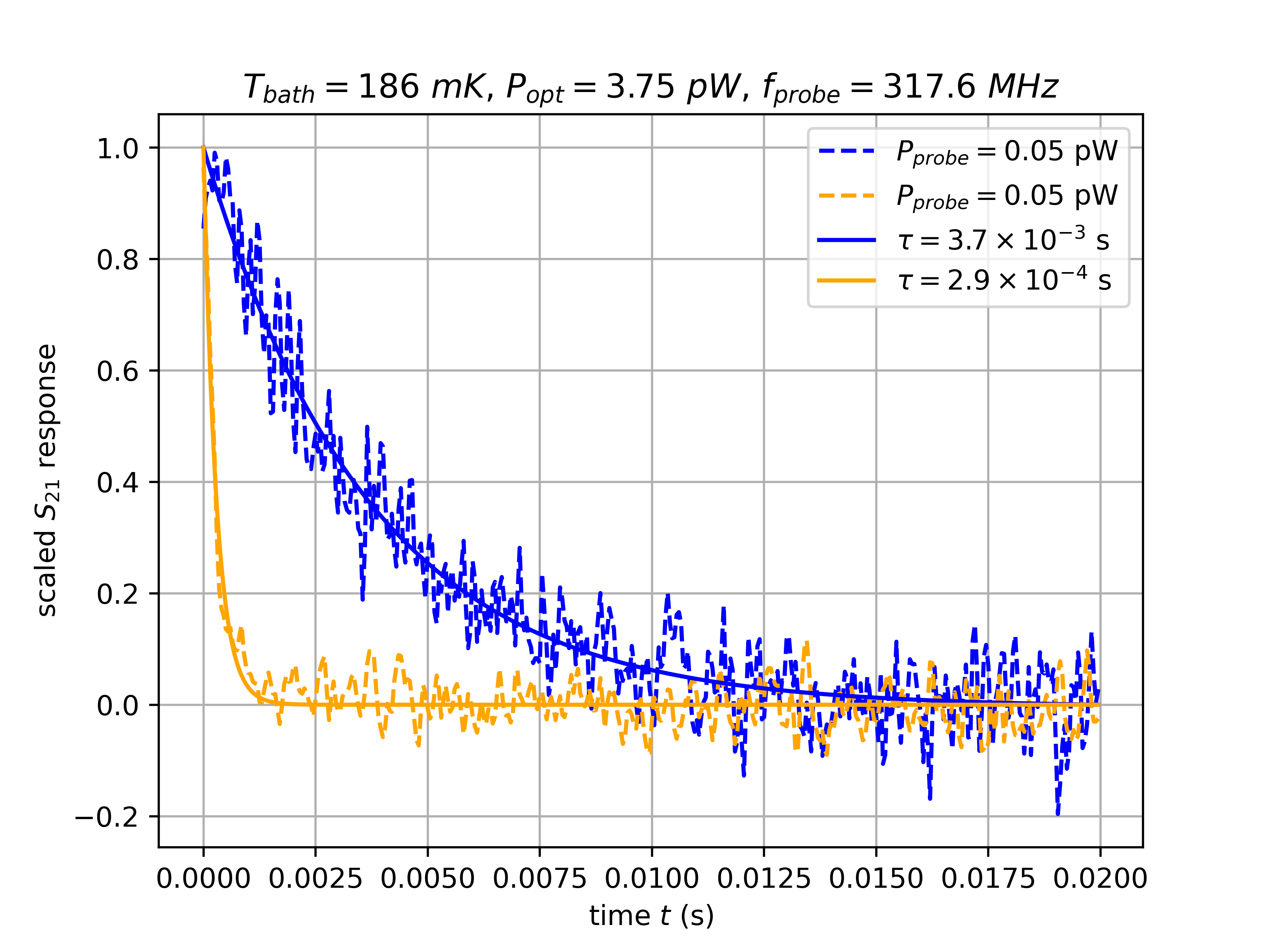}
\caption{\label{fig:timeconstants} Measured $S_{21}$ step responses to a small change in incident power at low (blue) and high (orange) probe power, averaged over $\approx 10$ periods shown in dotted lines. Solid lines show the fit to the exponential decay at this edge of the applied $P_{opt}$ square wave. For clarity, the data are scaled and shifted along the $y$-axis, such that the response range for both is $(0, 1]$. At high probe power, the time for $S_{21}$ to reach a steady state is reduced dramatically due to strong negative electrothermal feedback.}
\end{figure}

We measured bolometer time constants on the hot branch at probe powers ranging from 0.05 pW to 8 pW. We applied probe signals at frequencies $f$ between 317.0 MHz and 318.0 MHz spaced by 50 kHz to the detector with bath temperature regulated to $T_{bath}$ = 186 mK. The incident power pulses with a 34 pW square impulse several time constants wide (0.05 s) to bias the detector into the hot branch. After the pulse, the incident power oscillates in a 1.92 Hz square wave between 3.65 and 3.85 pW. We recorded $S_{21}$ for 5 s intervals at a sample rate of 20 kHz. Stacked $S_{21}$ traces at low and high negative electrothermal feedback are shown in Fig. \ref{fig:timeconstants}. Fitting the exponential decay of detector response $S_{21}$ at the edge of the applied square wave estimates the time constant $\tau$. This gives us the loop gain $\mathcal L$ by Eq. \ref{eqn:loopgain}.

\begin{figure}[!ht]
\includegraphics[width=\columnwidth]{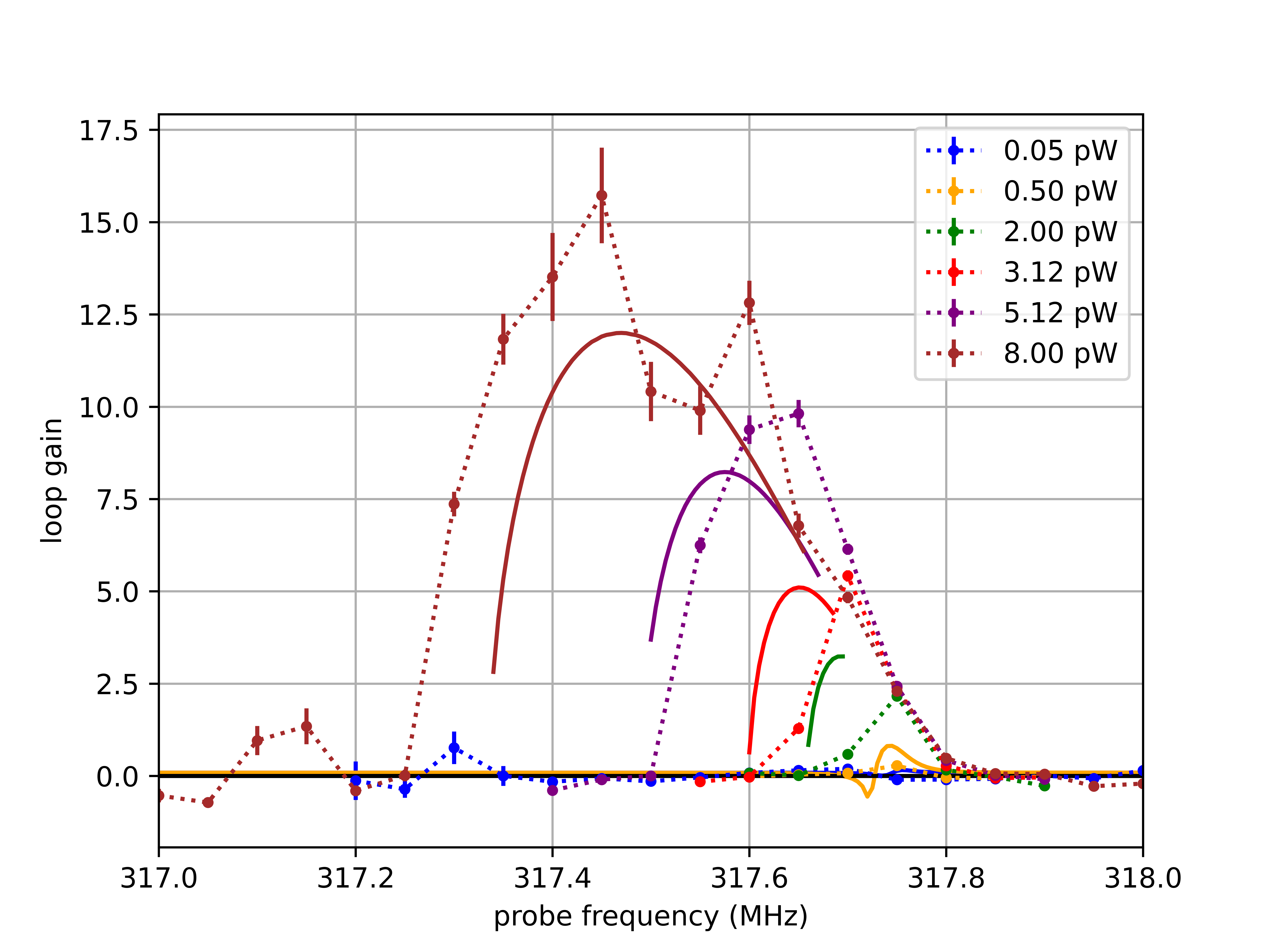}
\caption{\label{fig:loopgain} Measured loop gains plotted against probe frequency (in MHz) at incident power $P_{opt} =$ 3.75 pW and a range of probe powers $P_{probe}$. Measured data points are connected by dotted lines; solid lines give model predictions between the two bifurcation frequencies. Higher values of $\mathcal L$ are obtained at higher probe powers in good agreement with predictions.}
\end{figure}

We show the observed loop gain as a function of probe frequency and power, along with our model predictions, in Fig. \ref{fig:loopgain}. The highest value $\mathcal L$ = 15.7 $\pm$ 1.3 observed was at $P_{probe}$ = 8 pW and $f$ = 317.45 MHz. Our model slightly underestimates the speed up of the bolometer time constant, which will be investigated in future work.

\subsection{\label{linearity}Linearity of Response}

\begin{figure}[htbp]
\includegraphics[width=\columnwidth]{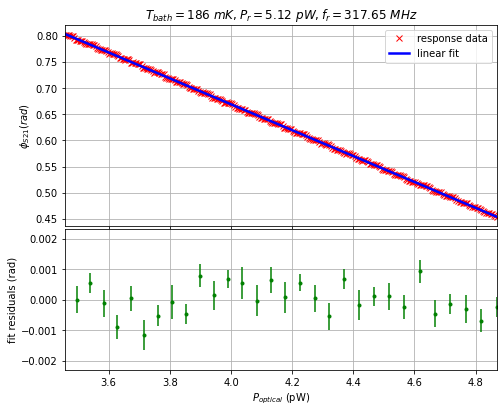}
\caption{\label{fig:linearity} Above, the response of the detector in $S_{21}$ phase over a range of incident power in red, along with a linear fit in blue. Below, the binned residual difference between data and linear fit, giving a reduced $\chi^2$ $\approx$ 2.}
\end{figure}

To measure linearity in the strong negative electrothermal feedback regime, we biased the TKID with $P_{probe}$ = 5.12 pW, $f$ = 317.65 MHz while regulating the bath temperature to $T_{bath}$ = 186 mK, where loop gain was previously measured to be $\mathcal L$ $\approx$ 10 at $P_{opt}$ = 3.75 pW. We swept incident power $P_{opt}$ over a wide range, from which we select a region from 3.5 pW to 4.9 pW (or 8\% below and 30\% above the nominal 3.75 pW) where the response is highly linear. The top panel of Fig. \ref{fig:linearity} shows the detector response to the incident power sweep in $S_{21}$ phase. We fitted this response to a linear model, as shown in the two panels of Fig. \ref{fig:linearity}. The residuals (lower panel) are dominated by measurement noise and indicate linearity better than 0.1\% in this operational range. 

\subsection{\label{nep} NEP Measurements}

\begin{figure}[!ht]
\includegraphics[width=\columnwidth]{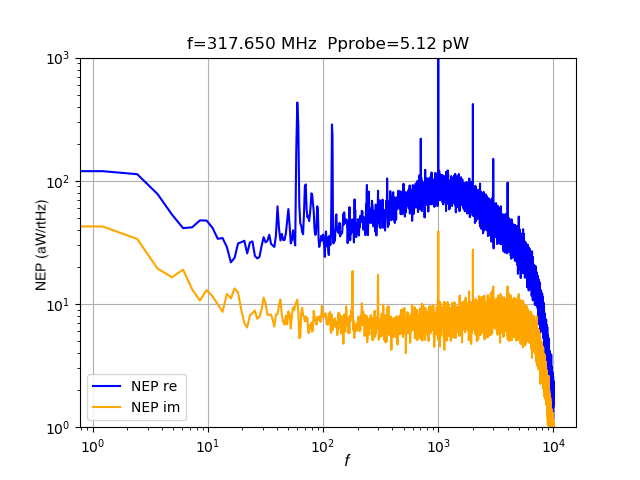}
\caption{\label{fig:nepspectrum} Noise-equivalent Power (NEP) amplitude spectrum for $P_{probe}$ = 5.12 pW, $f$ = 317.65 MHz and $P_{opt}$ = 3.75 pW on the hot branch. The blue curve shows thermal noise (in the direction in the complex plane that changing incident power changes $S_{21}$), while the yellow curve gives readout noise (in the orthogonal direction). }
\end{figure}

We measured the noise-equivalent power (NEP) of the TKID over a range of probe frequencies and powers. As in the time constant measurements, we biased the detector into the hot branch by pulsing the incident power after applying the probe signal. The NEP amplitude spectrum, shown in Fig. \ref{fig:nepspectrum} for $P_{probe}$ = 5.12 pW and $f$ = 317.65 MHz, rises at kHz frequencies, suggestive of internal thermal resistance and multiple decoupled heat capacities discussed in Section \ref{sec:noisemodel}.

We suspect common-mode environmental noise, likely from thermal fluctuations or RF-interference susceptibility that are detector-detector correlated, to be responsible for the higher noise amplitude between 10 - 30 Hz and the 1/$f$ knee at $\approx$ 10 Hz \cite{wandui_thermal_2020}. We did not implement a pair differencing scheme that would remove common-mode noise to obtain the results shown in Fig. \ref{fig:nepspectrum}. These methods have been previously proven effective on a small array of detectors, achieving phonon-limited NEPs with a 1/$f$ knee at $\approx$ 0.1 Hz for the same design \cite{wandui_thermal_2020}. 

\begin{figure}[htbp]
\includegraphics[width=\columnwidth]{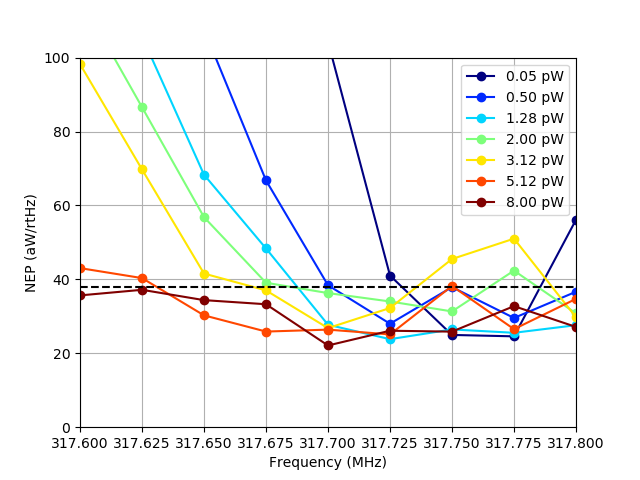}
\caption{\label{fig:nepall} NEP at several probe fixed signals $P_{probe}$ and $f$ on the negative electrothermal feedback hot branch.}
\end{figure}

Fig. \ref{fig:nepall} shows the NEP at all measured probe powers and frequencies. The NEP is averaged over 10 - 30 Hz to minimize the impact of the excess low-frequency noise and the rise in kHz high-frequency noise. There is a probe frequency at all probe powers that achieves a NEP of 25 aW/rtHz. This is below the intended photon noise of 38 aW/rtHz, but well above the expected phonon noise of 14 aW/rtHz. We believe that the NEP is limited by environmental noise, as we have observed phonon-limited noise with much lower 1/$f$ knees in a TKID of the same design in a different testbed \cite{wandui_thermal_2020}.

\section{\label{conclusion} Conclusion}
We have demonstrated negative electrothermal feedback in a TKID biased at readout probe powers comparable to the design incident power through the speed-up in bolometer response time. The maximum speed-up we observed was a factor of $\tau_0/\tau$ = 16.7 $\pm$ 1.3 at $P_{probe}$ = 8 pW and $P_{opt}$ = 3.75 pW. 

At a bias of $P_{probe}$ = 5.12 pW, we observed a highly linear response in phase $\phi_{S_{21}}$ for $P_{opt}$ = 3.5 - 4.9 pW. We also provide noise-equivalent power (NEP) measurements of the detector which demonstrate that the noise performance remains comparable to nominal TKID values when they are biased at low probe powers. Our NEP measurements do not reach the phonon noise floor, so we cannot yet rule out whether high probe power operation introduces additional noise at the 15 aW/rtHz level.

In this work, we operated only a single detector in the negative electrothermal feedback regime. As the TKID is intended to be a multiplexable detector, it remains to be seen what the practical implication of high probe power operation is on multiplexing. We anticipate an increase in the dynamic range requirement of the readout due to the decrease in responsivity relative to the probe power. Additionally, moving resonators with probe power can impact resonator collisions.

We see hints of the internal thermal structure to the detector in our NEP measurements, manifesting as a characteristic rise at high frequency. We intend to measure the incident power to $S_{21}$ transfer function in the frequency domain to clarify the internal thermal structure of the TKID. The results presented here indicate that operating the detector with high electrothermal feedback does not degrade the noise in single devices; we leave further noise performance tests using common-mode noise removal to future work.

The increase in speed due to negative electrothermal feedback could allow TKIDs to be practical in low incident power applications such as in narrow bandwidth line intensity mapping spectrometers, where low thermal conductance would lead to the TKID being otherwise too slow. The increase in speed could also be useful in calorimeters, where energy resolution is inversely proportional to the square root of the loop gain \cite{irwin_transition-edge_2005}.

\begin{acknowledgments}
Our work was carried out at the Jet Propulsion Laboratory, Caltech, under contract from the National Aeronautics and Space Administration. We thank Mark Lindeman and Jonas Zmuidzinas for insightful discussion. Shubh Agrawal's work was supported by the Dr. Gary Stupian SURF Fellowship.
\end{acknowledgments}

\section*{Data Availability}
The data that support the findings of this study are available from the corresponding author upon reasonable request.

\bibliography{references}

\end{document}